# Hollow Gaussian beam generation through nonlinear interaction of photons with orbital-angular-momemtum

N. Apurv Chaitanya,[1,2] M. V. Jabir,[1] J. Banerji,[1] G. K. Samanta[1]

[1]*Photonic Sciences Lab., Physical Research Laboratory, Ahmedabad 380009, Gujarat, India*
[2]*Indian Institute of Technology-Gandhinagar, Ahmedabad 382424, Gujarat, India*
*\*Corresponding author: apurv@prl.res.in*

**Hollow Gaussian beams (HGB) are a special class of doughnut shaped beams that do not carry orbital angular momentum (OAM). Such beams have a wide range of applications in many fields including atomic optics, bio-photonics, atmospheric science, and plasma physics. Till date, these beams have been generated using linear optical elements. Here, we show a new way of generating HGBs by three-wave mixing in a nonlinear crystal. Based on nonlinear interaction of photons having OAM and conservation of OAM in nonlinear processes, we experimentally generated ultrafast HGBs of order as high as 6 and power >180 mW at 355 nm. This generic concept can be extended to any wavelength, timescales (continuous-wave and ultrafast) and any orders. We show that the removal of azimuthal phase of vortices does not produce Gaussian beam. We also propose a new and only method to characterize the order of the HGBs.**

The dark hollow beams (DHB) are identified with their characteristic doughnut intensity distribution, a dark center enclosed by a bright ring in the beam cross section. Like conventional DHBs such as optical vortices[1], higher order Bessel[2] and Mathieu[3] beams, HGBs[4,5] also have doughnut intensity profile but do not carry any OAM. In addition to the vast applications ranging from atomic optics to plasma physics[6-15], HGBs have also attracted a great deal of scientific interest in understanding its propagation and transformation dynamics[5, 16, 17, 18]. Generation of HGBs have been realized using linear optical elements in different methods such as spatial filtering[19], geometrical optics[20], fibers[21], spatial-light-modulator[16], and Laguerre-Gaussian (LG) beam transformation[22]. However, nonlinear generation processes enable HGBs to have a new wavelength across electromagnetic spectrum and also to have high output power and higher order in all timescales as required for most of the applications[6-15].

The HGBs have a similar functional form[5] as that of optical vortex beams except the azimuthal phase term *exp(-ilθ)* where, *l* is the topological charge or OAM mode of the vortex. Therefore, HGBs can be generated by removing the azimuthal phase term of the vortices[22]. Given that the nonlinear frequency conversion processes[23-25] satisfy OAM conservation[26,27], one can in principle, remove the azimuthal phase terms of the generated beam through annihilation of OAM modes of the interacting beams in three wave-mixing process. As a proof of principle, here we report, for the first time to the best of our knowledge, nonlinear generation of HGBs. Based on sum frequency mixing of two OAM carrying ultrafast beams at 1064 nm and 532 nm having equal OAM orders but of opposite helicity in a nonlinear medium, we have generated HGBs of order as high as 6 and output power as much as 180 mW at 355 nm. It is a generic concept and can be extended to any wavelength and timescale. In addition, by controlling the sign of the helicity of OAM modes one can generate higher order optical vortices at desired wavelengths. We also propose a new and (at present,) only method to characterize the order of HGBs. On contrary to the common belief[28], the present study show that the removal of azimuthal phase of vortices does not produce Gaussian beam.

**Theory**
For theoretical understanding of nonlinear generation of HGBs we consider sum frequency generation (SFG) of two pump vortex beams (denoted by *m*=1, 2) with transverse electric field amplitude given as[1]

$$E_m(\rho, \phi) = \left(\frac{\rho}{w_G}\right)^{|l_m|} \exp\left(-\frac{\rho^2}{w_G^2}\right) \exp(il_m\phi), \qquad (1)$$

where, $l_m$ is the order of the vortex and $w_G$ is the waist radius of the Gaussian beam hosting the vortex. From the coupled wave equations of SFG process[23] under perfect-matching, the transverse electric field amplitude of the generated field can be represented in the form,

$$E_3(\rho, \phi) = \left(\frac{\rho}{w_o}\right)^{|l_1|+|l_2|} \exp\left(-\frac{\rho^2}{w_o^2}\right) \exp(i[l_1 + l_2]\phi). \qquad (2)$$

Here, $w_o$ is the waist radius of the generated beam. When the pump beams have same OAM orders but opposite helicity ($l_1$=-$l_2$=$l$, as schematically shown in Fig. 1a), the field amplitude of the generated beam (Eq. 2) will have the form of a HGB[5],

$$E = \left[\left(\frac{\rho}{w_o}\right)^2\right]^l \exp\left(-\frac{\rho^2}{w_o^2}\right). \qquad (3)$$

The order of HGB is same as the order *l* of the pump vortices. This simple treatment clearly indicates the possibility of nonlinear generation of HGBs with desired orders and frequency across the electromagnetic spectrum.

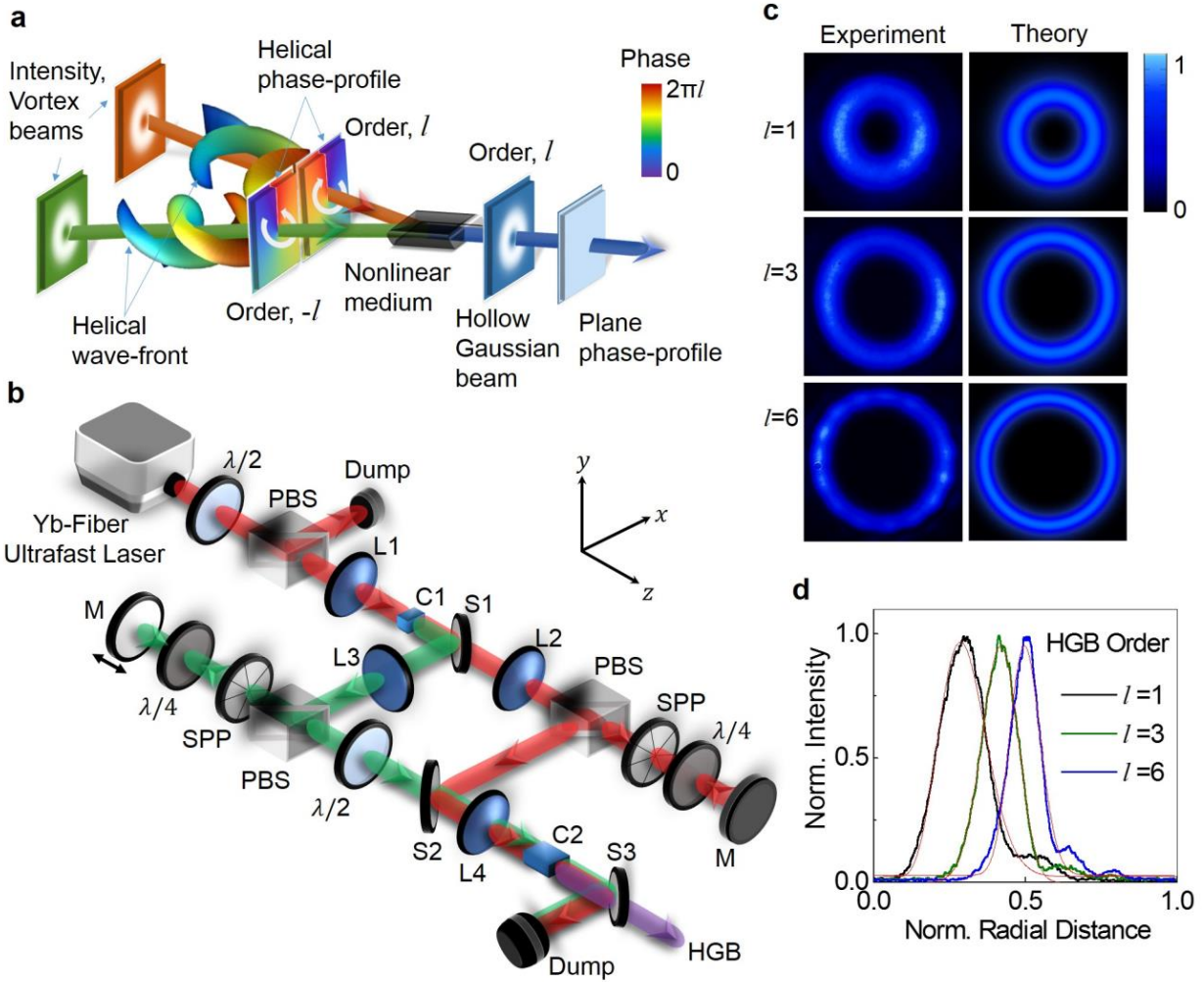

Fig. 1. **Nonlinear generation of hollow Gaussian beam**. **a**, Pictorial representation of SFG of optical vortex beams of same OAM orders but opposite helicity producing HGBs. **b**, Schematics of the experimental setup. $\lambda/2$, Half-wave-plate; PBS, Polarizing beam splitter cube; L1-4, Lenses; C1-2, Nonlinear crystals for wave-mixing; SPP, Spiral phase plate; S1-3, Dichroic mirrors; $\lambda/4$, Quarter-wave plate. **c,** Experimental (first column) and theoretical (second column) images of the intensity profile of HGBs of order, $l$= 1, 3 and 6 (from top to bottom).**d**, Line intensity profile of the generated HGBs of order, $l$= 1, 3 and 6.

**Results and discussions**

The schematics of the experimental setup is shown in Fig. 1b (also see Methods). A Yb-fiber ultrafast laser with Gaussian intensity profile at 1064 nm is frequency-doubled into green at 532 nm[25]. Using spiral phase-plates (SPP) and vortex-doubler[27] both the unconverted beam at 1064 nm and green beam at 532 nm are transformed into vortices of order up to $l=\pm6$. The direction of phase variation (clock-wise or anti-clock wise) of SPP to the input beam determines the sign of its vortex. By careful choice of the beam direction with respect to SPP, we can make the pump beams at 1064 nm and 532 nm to carry vortices of same order but opposite signs. These beams, while interacting in bismuth-borate (BIBO) crystal through SFG process, produce optical radiation at 355 nm. First column of Fig. 1c shows experimental intensity profile of the SFG beam for pump vortices of order $l$=1, 3, 6. Like vortex beams[27], the newly generated beams also carry dark core whose radius increases with the order. The theoretical intensity profiles (see second column of Fig. 1c), obtained from the coupled amplitude equations of SFG process with the experimental parameters, show a close agreement with the experiment.

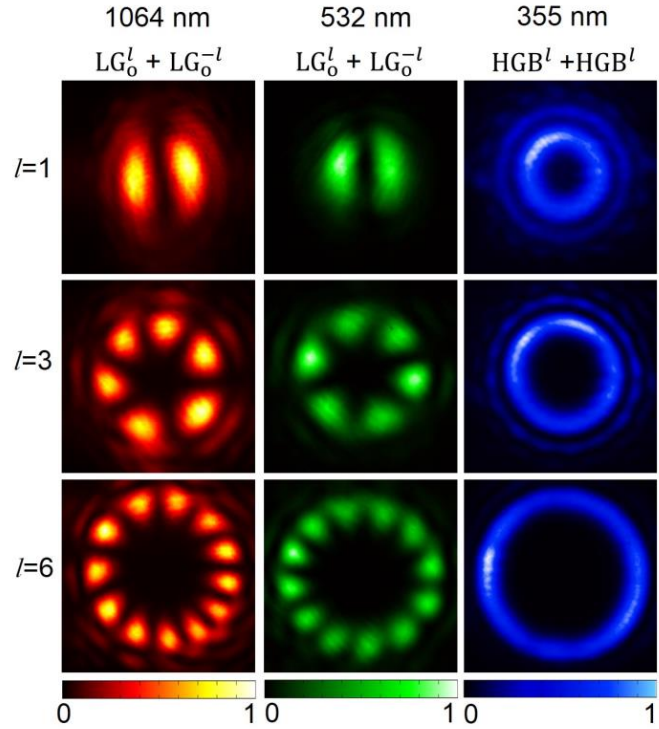

Fig. 2. **Verification of OAM modes of the interacting beams in SFG process for the confirmation of hollow Gaussian beam generation**. Experimental ring lattice intensity pattern of diagonal projection of superposition state, $\frac{1}{\sqrt{2}}(|H, LG^l\rangle + |V, LG^{-l}\rangle)$ of pump beams at 1064 nm (first column) and 532 nm (second column) confirming the existence of OAM in both the pump beams with order of $l$=1, 3, and 6 (top to bottom). Third column shows similar diagonal projection for the SFG beam confirming nonexistence of azimuthal phase (OAM) and successful generation of HGBs of order $l$=1, 3, and 6 (top to bottom).

The striking similarity between the line intensity profiles (see Fig. 1d) of the SFG beams to that of the characteristic intensity line profile[5] of HGB indicates the possibility of generation of HGB through nonlinear process. However, the vortex beams also have intensity profiles similar to that of HGBs but carry OAM. Therefore, to confirm the generation of HBGs through nonlinear interaction, we verified the absence of OAM in SFG beam generated from pump beams having OAM using inteferometric technique[29]. Each of the beams is passed through a balanced polarizing beam-splitter (PBS) cube based Mach-Zehnder interferometer (MZI)[29] with a dove prism in one of its arm and recorded the beam (see Fig. 2) after a polarizer (see Methods). The first and second column of Fig. 2, show the ring lattice structure (a signature pattern of azimuthal phase variation) and corresponding number of lobes confirms the vortex order of both the pumps to be $l$=1, 3 and 6. However, the absence of ring structure in the superposition state of SFG beams (see third column of Fig. 2), a clear indication of nonexistence of azimuthal phase variation and OAM, confirms nonlinear generation of HGBs. The annihilation of OAMs of the interacting pump beams in nonlinear processes results in HGBs with zero OAM.

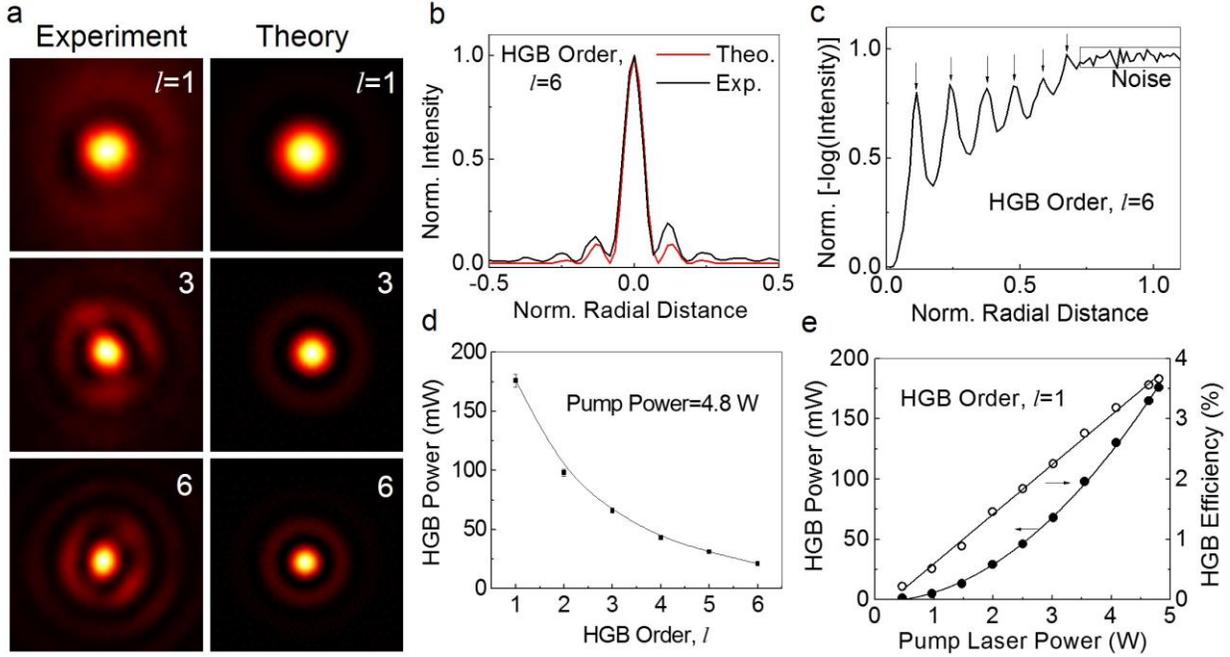

Fig. 3. **Determination of order of HGBs and characterization of the nonlinear generation of HGBs generation**. **a**, Experimental (first column) and theoretical (second column) images of intensity profile of HGBs, of orders, $l$=1, 3 and 6 (top to bottom) recorded at Fourier plane. **b**, Experimental (red line) and theoretical (black line) radial intensity distribution of HGB of order $l$=6 at the Fourier plane. **c**, Normalized negative logarithmic radial intensity distribution of the HGB recorded at the Fourier plane reveals the order of HGB (here order is $l$=6) **d**, Exponential decay in the output power of HGBs with its order. **e**, Quadratic (linear) dependence of HGB power (efficiency) to the input laser power.

None of the existing techniques used to reveal the order of DHBs can be extended to determine the order of HGBs. Therefore, it is imperative to devise a new technique to determine the order of HGBs. In polar co-ordinates, the intensity distribution of the Fourier transformation (FT) of HGB of order, $l$ has the form[5]

$$\tilde{I}(\rho) = |exp(-\rho^2)\mathcal{L}_l(\rho^2)|^2 \qquad (4)$$

Here, $\mathcal{L}$ is the Laguerre polynomial with radial index, $l$. The intensity distribution of FT of HGB [see Eq. (4)] of order $l$, is a Laguerre-Gaussian beam (radial index, $l\neq 0$ and azimuthal index, $p$=0) modulated by a Gaussian envelope. Unlike the intensity distribution of FT of vortex beam resulting in a vortex beam with a dark hole at the center, the intensity distribution of FT of HGBs produces maximum intensity point at the center ($\rho$=0) followed by ripples corresponding to the order, $l$ of HGBs. Counting the number of ripples one can, in principle, determine the order of

HGBs. In our experiment we recorded the intensity distribution of the HGB of orders $l$=1, 3 and 6 (first column of Fig. 3a) at the Fourier plane of a lens of focal length $f$=150 mm and found that the number of rings are increasing with the order $l$. The theoretical images (second column of Fig. 3a) obtained from Eq. (4) using experimental parameters show close agreement with the experimental images. For a better understanding, we recorded the theoretical (red line) and experimental (black line) line profiles for FT of HGB of order $l$=6 (Fig. 3b). However, it is difficult to appreciate the number of ripples present in the images in order to find the order of the HGBs. This is due to that fact that the exponential decay term present in the intensity distribution of FT of the HGB [see Eq. (4)] falls steeply as compared to the Laguerre polynomial and suppresses the characteristic ripple structures. For accurate determination of the order of HGBs we have changed intensity scale from linear to negative log-scale (see Fig. 3c) enhancing the visibility of the small details of intensity distribution of FT of HGBs. The HGB of order $l$=6 have clear six peaks (arrows in Fig. 3c) confirming the reliability of our technique in determining the order of HGBs.

We studied the variation of HGB output power with its order (Fig. 3d). For a fixed fundamental pump power (~4.6 W), we found that the HGB power decreases exponentially from ~180 mW to 21 mW with the increase of its order from $l$=1 to 6. Such drop in HGB power can be attributed to the decrease of nonlinear gain with the order of the pump vortex similar to the effect observed in nonlinear generation of vortex beams[27]. We also observed the power scaling effect in the nonlinear generation of HGB (Fig. 3e) of order $l$=1. The output power (efficiency) of the HGB shows quadratic (linear) dependence to laser power in SFG process[24] resulting in maximum power of 180 mW with single-pass IR-UV conversion efficiency of 3.7%.

In conclusion, we have demonstrated, for the first time to the best of our knowledge, that annihilation of two optical vortices of same order with opposite sign in a nonlinear medium produces a new class of dark core beam, known as HGB. We have generated HGBs of order as high as 6 and power as high as 180 mW at 355 nm. The nonlinear generation schemes can be used to generate HGB at any wavelength across the electromagnetic spectrum. We have also devised a new and only method to characterize the order of such HGBs.

**Methods**

**Experimental setup**: The schematic of the experimental setup for nonlinear generation of HGBs is shown in Fig. 1b. An ultrafast Yb- fiber laser of average power 5 W is used as the primary laser source. The laser produces femtosecond pulses in *sech²* shape with temporal width (full width at half-maximum, FWHM) of ~260 *f*s at a repetition rate of 78 MHz. The spectral width (FWHM) of the laser is measured to be ~15 nm centered at 1060 nm. Operating the laser at its highest power to access its optimum performance in terms of temporal width, a combination of half wave plate ($\lambda/2$) and a polarization beam splitter (PBS) cube is used to control the laser power to the nonlinear crystals. A lens (L1) of focal length $f$=50 mm is used to focus the beam at the center of a 1.2-mm-long BIBO ($BiB_3O_6$) crystal (C1) for second harmonic generation (SHG) of Yb-fiber laser into green at 532 nm. The SHG stage is similar to the one used in our previous work[24]. The SHG stage produces 1.7 W of green power for fundamental power of 4.6 W in Gaussian beam profile. A dichroic mirror (S1) is used to extract green radiation from the fundamental wavelength. Lenses, L2 and L3, both of focal length ($f$=100 mm), collimate unconverted fundamental and generated green beams respectively. The green beam and unconverted fundamental beam are used as pump beams for sum frequency generation (SFG) process. Using two spiral phase plates, SPP1 and SPP2, having spiral winding corresponding to vortex order of 1 and 2 respectively and a vortex-doubler setup[27] comprising of a PBS, quarter wave-plate ($\lambda/4$) and a high reflectance mirror (M), we separately transformed the green and unconverted fundamental radiation in Gaussian spatial profiles into optical vortices of order, $l$=1-6. The working principle of the vortex-doubler can be found elsewhere[27]. The direction of phase variation (clock-wise or anti-clock wise) of SPP to the input beam determines the sign ($\pm l$) of its vortex. By careful choice of the beam direction with respect to SPP, we can make both green and unconverted fundamental beams to carry vortices of the same order but with same/opposite sign. These two beams are recombined using a dichroic mirror (S2) and focused at the center of a 5-mm-long $BiB_3O_6$ (BIBO) crystal (C2) cut at $\theta$=137° ($\varphi$=90°) in $yz$ optical plane for collinear type–I ($e + e \rightarrow o$) phase-matched SFG of 1064 nm and 532 nm at normal incidence. One of the mirrors (M) is placed on a translation stage for temporal overlapping of the interacting beams in the nonlinear crystal (C2). A dichroic mirror (S3) is used to extract the SFG beam at 355 nm from the unconverted beams at 1064 nm and 532 nm. The intensity profile of the beam is recorded using a CCD based beam profiler.

**Confirmation of nonlinear generation of HGBs**:

The azimuthal phase variation of beams, a signature of OAM modes, is measured using an interferometric technique[29]. An optical beam having OAM (LG, Laguerre-Gaussian) mode of arbitrary order $l$ polarized at 45° (diagonal) can be represented as $|D, LG^l\rangle = \frac{1}{\sqrt{2}}(|H, LG^l\rangle + |V, LG^l\rangle)$, where $|H\rangle$ and $|V\rangle$ denote the orthogonal polarisation states. If such a beam passes through balanced polarising beam splitter based Mach-Zehnder interferometer (MZI)[29] with a Dove prism in one of its arm, the extra reflection due to Dove prism transforms the input beam into a new superposition state represented as $\frac{1}{\sqrt{2}}(|H, LG^l\rangle + |V, LG^{-l}\rangle)$. The diagonal projection of this state having the form $\frac{1}{\sqrt{2}}(|LG^l\rangle + |LG^{-l}\rangle)$ produces a ring lattice structure with $2l$ number of radial fringes or petals. Therefore, 2, 6 and 12 number of petals in the first two columns of Fig. 2 confirms that both the pump beams at 1064 nm (first column of Fig. 2) and 532 nm (second column of Fig. 2) have OAM modes $l$=1, 3 and 6. However, due to the nonexistence of OAM mode in HGBs, the extra reflection due to Dove prism does not alter the output state from the input state, $\frac{1}{\sqrt{2}}(|H, HGB^l\rangle + |V, HGB^l\rangle)$. Here, hollow Gaussian beam mode of order, $l$ is represented by $|HGB^l\rangle$. The diagonal projection of such state produces output beam of the form $|HGB^l\rangle$, the HGB itself (see third column of Fig. 2). Such observation proves the nonexistence of azimuthal phase and OAM and successful generation of HGBs of order up to 6.